\documentclass[a4paper,11pt]{article}
\usepackage{pos}

\usepackage{tikz}
\usepackage[compat=1.1.0]{tikz-feynman}
\usetikzlibrary{arrows,matrix,positioning,fit,calc}

\usepackage{comment}

\usepackage[normalem]{ulem}

\title{
Dineutrino modes probing lepton flavor violation
}

\author[a]{Rigo Bause}
\author*[a]{Hector Gisbert}
\author[a]{Marcel Golz}
\author[a]{Gudrun Hiller}

\affiliation[a]{Fakult\"at f\"ur Physik, TU Dortmund,\\ Otto-Hahn-Str.\,4, D-44221 Dortmund, Germany}

\emailAdd{rigo.bause@tu-dortmund.de}
\emailAdd{hector.gisbert@tu-dortmund.de}
\emailAdd{marcel.golz@tu-dortmund.de}
\emailAdd{ghiller@physik.uni-dortmund.de}

\abstract{$SU(2)_L$--invariance links charged dilepton $\bar q\,q^\prime\,\ell^+\,\ell^-$ and dineutrino $\bar q\, q^\prime\, \bar\nu\,\nu$ couplings. 
This connection can be established using the Standard Model Effective Field Theory framework, and allows to perform complementary experimental tests of lepton universality and charged lepton flavor conservation with flavor-summed dineutrino observables. 
We present its phenomenological implications for the branching ratios of rare charm decays $c\to u\,\nu\,\bar\nu$ and rare $B$ decays $b\to s\,\bar\nu\,\nu$ decays.}

\FullConference{%
  The European Physical Society Conference on High Energy Physics (EPS-HEP2021),\\
  26-30 July 2021\\
  Online conference, jointly organized by Universität Hamburg and the research center DESY
}

\begin{document}
\maketitle

\section{Introduction}

Flavor-changing neutral currents (FCNCs) of $q^\alpha$ and $q^\beta$ quarks induced by $|\Delta \,q^\alpha|=|\Delta\, q^\beta|=1$ processes represent excellent probes of New Physics (NP) beyond the Standard Model (SM). Their weak loop suppression triggered by the Glashow-Iliopoulos-Maiani (GIM) mechanism and  Cabibbo-Kobayashi-Maskawa (CKM) hierarchies, not necessarily present in SM extensions, can result in large experimental deviations from the SM predictions alluding to a breakdown of SM symmetries. In addition, its environment is enriched with further tests if leptons are involved, that is $q_\alpha q_\beta\,\ell_i^+\ell_j^-$ and $q_\alpha q_\beta\,\bar\nu_i\nu_j$. We exploit the $SU(2)_L$-link between left-handed charged lepton and neutrino couplings, which may be used to assess charged lepton flavor conservation (cLFC) and lepton universality (LU) quantitatively using flavor-summed dineutrino observables~\cite{Bause:2020auq}. This link~\eqref{eq:super} is presented for $|\Delta\, q^\alpha|=|\Delta\, q^\beta|=1$ processes, but we stress that it holds analogously for other conserved quark transitions, both in the up- and down-sector. 

These proceedings are organized as follows: In Section~\ref{sec:EFT}, we present the effective theory framework where the $SU(2)_L$--link between neutrino and charged lepton couplings is derived. In Sections~\ref{sec:charm} and \ref{sec:beauty}, we work out its phenomenological implications for charm and beauty, respectively. The conclusions are drawn in Section~\ref{sec:con}. The results are based on Refs.~\cite{Bause:2020auq,Bause:2020xzj,Bause:2021ply}, we refer there for further details.

\section{$\boldsymbol{SU(2)_L}$--link between dineutrino and charged dilepton couplings}\label{sec:EFT}

At lowest order in the SM effective field theory (SMEFT), the Lagrangian accounting for semileptonic (axial-)vector four-fermion operators is given by~\cite{Grzadkowski:2010es}, 
\begin{align}\label{eq:ops} 
{\mathcal{L}}_{\text{eff}} & \supset \frac{C^{(1)}_{\ell q}}{v^2} \bar Q \gamma_\mu Q \,\bar L \gamma^\mu L +\frac{C^{(3)}_{\ell q}}{v^2} 
\bar Q \gamma_\mu  \tau^a Q \,\bar L \gamma^\mu \tau^a L   +\frac{C_{\ell u}}{v^2}  \bar U \gamma_\mu U \,\bar L \gamma^\mu L +
\frac{C_{\ell d }}{v^2} \bar D \gamma_\mu D \,\bar L \gamma^\mu L \,. 
\end{align}
Reading off couplings to dineutrinos $(C_A^N)$ and charged dileptons $(K_A^N)$ by writing the operators~\eqref{eq:ops} into $SU(2)_L$-components, one obtains
\begin{align}\label{eq:links} 
\begin{split}
C_L^U&=K_L^D={\frac{2\pi}{\alpha}}\left(C^{(1)}_{\ell q} + C^{(3)}_{\ell q}\right)  \,, \, C_R^U=K_R^U={\frac{2\pi}{\alpha}}C_{\ell u}  \,, \\
C_L^D&=K_L^U={\frac{2\pi}{\alpha}}\left(C^{(1)}_{\ell q}  - C^{(3)}_{\ell q}\right)  \,, \,
C_R^D=K_R^D={\frac{2\pi}{\alpha}}C_{\ell d}  \,,
\end{split}
\end{align}
where $N=U$ ($N=D$) represents the up-quark sector (down-quark sector), and $A=L (R)$ denotes left- (right-) handed quark currents. Interestingly, $C_R^N=K_R^N$ holds model-independently, while $C_L^N$ is not fixed by $K_L^N$ in general due to the different relative signs of $C^{(1)}_{\ell q}$ and $C^{(3)}_{\ell q}$. Expressing Eqs.~\eqref{eq:links} in the mass basis, that is $\mathcal{C}_{L}^{N} = W^\dagger  \,\mathcal{K}_{L}^{M}\, W+ {\mathcal{O}}(\lambda)$, $\mathcal{C}_{R}^{N}=W^\dagger \,\mathcal{K}_{R}^{N}\, W$ where $W$ is the Pontecorvo-Maki-Nakagawa-Sakata (PMNS) matrix and $\lambda \sim 0.2$ the Wolfenstein parameter,  
and summing lepton flavors $i,j$ incoherently, one obtains the following identity~\cite{Bause:2020auq}
\begin{align}
   &\sum_{\nu=i,j} \left( \vert\mathcal{C}_L^{{N} ij}\big\vert^2+\vert\mathcal{C}_R^{{N} ij}\big\vert^2 \right)=\sum_{\ell=i,j} \left(   \vert\mathcal{K}_L^{{M} ij}\big\vert^2+\vert\mathcal{K}_R^{{N} ij}\big\vert^2 \right)  + {\mathcal{O}}(\lambda)~,\label{eq:super} 
\end{align}
between charged lepton couplings $\mathcal{K}_{L,R}$ and neutrino ones $\mathcal{C}_{L,R}$.\footnote{Wilson coefficients in calligraphic style denote those for mass eigenstates.} 
Here, we use $N,M=U,D$ when the link is exploited for neutrino couplings in the up-quark sector, while $N,M=D,U$ for the down-quark sector. Eq.~\eqref{eq:super} allows the prediction of dineutrino rates for different leptonic flavor structures $\mathcal{K}_{L,R}^{N \, ij}$,
\begin{itemize}
\item[{\it i)}] $\mathcal{K}_{L,R}^{N \, ij} \propto \delta_{ij}$,  \textit{i.e.} \emph{lepton-universality} (LU),
\item[{\it ii)}] $\mathcal{K}_{L,R}^{N \, ij}$ diagonal, \textit{i.e.}  \emph{charged lepton flavor conservation} (cLFC),
\item[{\it iii)}] $\mathcal{K}_{L,R}^{N \, ij}$ arbitrary,
\end{itemize}
which can be probed with lepton-specific measurements. In the following sections, we use the following notation \textit{i.e.}~$\mathcal{K}_{L,R}^{bs{ij}}\,=\mathcal{K}_{L,R}^{D_{23}{ij}}$,~ $\mathcal{C}_{L,R}^{bsij}=\mathcal{C}_{L,R}^{D_{23}ij} $, etc., to improve the readability.

\addtolength{\tabcolsep}{2pt} 
\begin{table}[h!]
    \centering
    \begin{tabular}{c|cccccc}
    \hline
    \hline
     & $ee$ & $\mu \mu $ & $\tau \tau$  & $e \mu$ & $e \tau$ & $\mu \tau$\\
    \hline
     $R^{\ell \ell'}$& $21$ & $6.0$ & $77$ & $6.6$ & $59$ & $70$ \\
     $\delta R^{\ell \ell'}$& $19$ & $5.4$ & $69$ & $5.7$ & $55$ & $63$ \\
     \hline
     $r^{\ell \ell'}$& $39$ & $11$ & $145$ & $12$ & $115$ & $133$ \\
     \hline
     \hline
    \end{tabular}
    \caption{Bounds on $|\Delta c|=|\Delta u|=1$ parameters $R^{\ell \ell'}$ and $\delta R^{\ell \ell'}$ from Eqs.~\eqref{eq:Rell}, as well as their sum, $r^{\ell \ell'}=R^{\ell \ell'}+\delta R^{\ell \ell'}$. Table taken from Ref.~\cite{Bause:2020xzj}.}
    \label{tab:limitsonR}
\end{table}
\addtolength{\tabcolsep}{-2pt}

\section{Predictions for charm}\label{sec:charm}

In this section, we study the implications of \eqref{eq:super} for $c \to u \,\nu \bar \nu$  dineutrino transitions, where the situation is exceptional as the SM amplitude is fully negligible due to an efficient GIM-suppression~\cite{Burdman:2001tf} and the current lack of experimental constraints. We use upper limits on $\mathcal{K}_{A}^{N\ell\ell^\prime}$ from high--$p_T$ \cite{Fuentes-Martin:2020lea,Angelescu:2020uug}, which allow to set constraints on 
\begin{align} 
 R^{\ell \ell'}&= \! |\mathcal{K}_L^{ sd \ell\ell'}|^2+ |\mathcal{K}_R^{cu \ell\ell'}|^2 \,,\qquad R_\pm^{\ell \ell'}=\! |\mathcal{K}_L^{sd \ell\ell'} \pm \mathcal{K}_R^{cu \ell\ell'}|^2 ,\,\label{eq:Rell}\\
 {\delta R^{\ell \ell'}}&=\!{ 2\,\lambda\,\text{Re}\left\lbrace\mathcal{K}_L^{ sd \ell\ell'}{\mathcal{K}_L^{ ss \ell\ell'}}^*- \mathcal{K}_L^{ sd \ell\ell'}{\mathcal{K}_L^{ dd \ell\ell'}}^*\right\rbrace ,}\nonumber
\end{align}
which directly enter in $c \to u \,\nu \bar \nu$ branching ratios. Upper limits on $R^{\ell \ell'}$, $\delta R^{\ell\ell^\prime}$ and their sum $r^{\ell \ell'}=R^{\ell \ell'}+\delta R^{\ell \ell'}$ are provided in Table~\ref{tab:limitsonR}. Since the neutrino flavors are not tagged, the  branching ratio is obtained by an incoherent sum
\begin{align} 
\mathcal{B}\left( c \to u \,\nu \bar \nu\right)=\sum_{i,j} \mathcal{B}\left( c\to u \,\nu_i  \bar \nu_j\right)\propto x_{uc}\,,
\end{align}
where $x_{uc}=\sum_{i,j}\left(|\mathcal C_L^{U ij}|^2+ |\mathcal C_R^{U ij}|^2\right)$. Using \eqref{eq:super} with $N,M=U,D$ and Table \ref{tab:limitsonR}, we obtain upper limits for the different benchmarks {\it i)}-{\it iii)}:
\begin{align}  \label{eq:LU}
x_{uc} &= 3\, r^{\mu \mu} \lesssim 34 \,, \quad  (\text{LU}) \\ \label{eq:cLFC}
x_{uc} &= r^{e e}\hspace{-0.1cm}+ r^{\mu \mu }\hspace{-0.1cm}+r^{\tau \tau}  \lesssim 196 \,, \quad (\text{cLFC}) \\ \label{eq:total}
x_{uc} &= r^{ee} \hspace{-0.1cm}+ r^{\mu \mu}\hspace{-0.1cm} +r^{\tau \tau} \hspace{-0.1cm}+2 \,( r^{e \mu}+ r^{e \tau}+r^{ \mu \tau} ) \lesssim 716 \,.
\end{align}
Since dimuon bounds are the most stringent ones, see Table \ref{tab:limitsonR}, they set the LU-limit \eqref{eq:LU}. Experimental measurements above the upper limit in \eqref{eq:LU} would indicate a breakdown of LU, while values above the limit in \eqref{eq:cLFC} would imply a violation of cLFC. Corresponding upper limits on branching ratios of dineutrino modes of a charmed hadron $h_c$ into a final hadronic state $F$, 
\begin{align}\label{eq:BR-A}
    \mathcal{B}(h_c\to F\,\nu \bar \nu)= A_+^{h_c F}\, x^+_{cu} + A_-^{h_c F} \, x^-_{cu} \,,
\end{align}
are provided in Table~\ref{tab:afactors} for several decays modes. The $A_\pm^{h_c F}$ coefficients in Eq.~\eqref{eq:BR-A} are given in the second column of Table~\ref{tab:afactors}. Using the limits \eqref{eq:LU}, \eqref{eq:cLFC}, \eqref{eq:total}, together with Eq.~\eqref{eq:BR-A} and the values of $A_\pm^{h_c F}$, we obtain upper limits on branching ratios for the three flavor scenarios  $\mathcal{B}_{\text{LU}}^{\text{max}}$, $\mathcal{B}_{\text{cLFC}}^{\text{max}}$, and  $\mathcal{B}^{\text{max}}$. A branching ratio measurement $\mathcal{B}_{\text{exp}}$ within $\mathcal{B}_{\text{LU}}^{\text{max}} < \mathcal{B}_{\text{exp}} < \mathcal{B}_{\text{cLFC}}^{\text{max}}$
would be a clear signal of LU violation. In contrast, a branching ratio above $\mathcal{B}_{\text{cLFC}}^{\text{max}}$ would imply a breakdown of cLFC.
\begin{table}[ht!]
 \centering
    \resizebox{0.45\textwidth}{!}{
  \begin{tabular}{lccccc}
  \hline
  \hline
  $h_c\to F $ & $A^{h_c\,F}_+$ & $A^{h_c\,F}_-$ &  $\mathcal{B}_{\text{LU}}^\text{max}$ & $\mathcal{B}_{\text{cLFC}}^\text{max}$ & $\mathcal{B}^\text{max}$ \\
  &$[10^{-8}]$&$[10^{-8}]$& $[10^{-7}]$ &$[10^{-6}]$ & $[10^{-6}]$\\
  \hline
  $D^0\to\pi^0$ & $0.9$ & -- & $ {6.1}$ & $ {3.5}$ & $ {13}$ \\
  $D^+\to\pi^+$ & $3.6$ & -- & $ {25}$ & $ {14}$ & $ {52}$ \\
  $D_s^+\to K^+$ & $0.7$ & --  & $ {4.6}$ & $ {2.6}$ & $ {9.6}$ \\
  &&&&   & \\
  $D^0\to\pi^0\pi^0$ & $\mathcal{O}(10^{-3})$ & $0.21$ & $ {1.5}$ & $ {0.8}$ & $ {3.1}$ \\
  $D^0\to\pi^+\pi^-$ & $\mathcal{O}(10^{-3})$ & $0.41$ & $ {2.8}$ & $ {1.6}$ & $ {5.9}$ \\
  $D^0\to K^+K^-$ & $\mathcal{O}(10^{-6})$ &  $0.004$ & $ {0.03}$ & $ {0.02}$ & $ {0.06}$ \\
  &&&& & \\
  $\Lambda_c^+\to p^+$ & $1.0$ & $1.7$ & $ {18}$ & $ {11}$ & $ {39}$ \\
  $\Xi_c^+\to \Sigma^+$ & $1.8$ & $3.5$ & $ {36}$ & $ {21}$ & $ {76}$ \\ 
  &&&&   &\\
  $D^0\to X$ & $2.2$ & $2.2$ & $ {15}$ & $ {8.7}$ & $ {32}$ \\
  $D^+\to X$ & $5.6$ & $5.6$ & $ {38}$ & $ {22}$ & $ {80}$ \\
  $D_s^+\to X$ & $2.7$ & $2.7$ & $ {18}$ & $ {10}$ & $ {38}$ \\
  \hline
  \hline
   \end{tabular}}
\caption{Coefficients $A^{h_c\,F}_\pm$, as defined in \eqref{eq:BR-A}, and model-independent upper limits on $\mathcal{B}_{\text{LU}}^{\text{max}}$, $\mathcal{B}_{\text{cLFC}}^{\text{max}}$, $\mathcal{B}^{\text{max}}$ from \eqref{eq:LU}, \eqref{eq:cLFC} and \eqref{eq:total}, respectively, corresponding to  the lepton flavor symmetry benchmarks {\it i)}-{\it iii)}. Table taken from Ref.~\cite{Bause:2020xzj}.
}
\label{tab:afactors}
\end{table}

\section{Testing lepton universality with $\boldsymbol{b\to s\,\nu\bar\nu}$}\label{sec:beauty}

\begin{figure}[h!]
    \centering
    \includegraphics[scale=0.55]{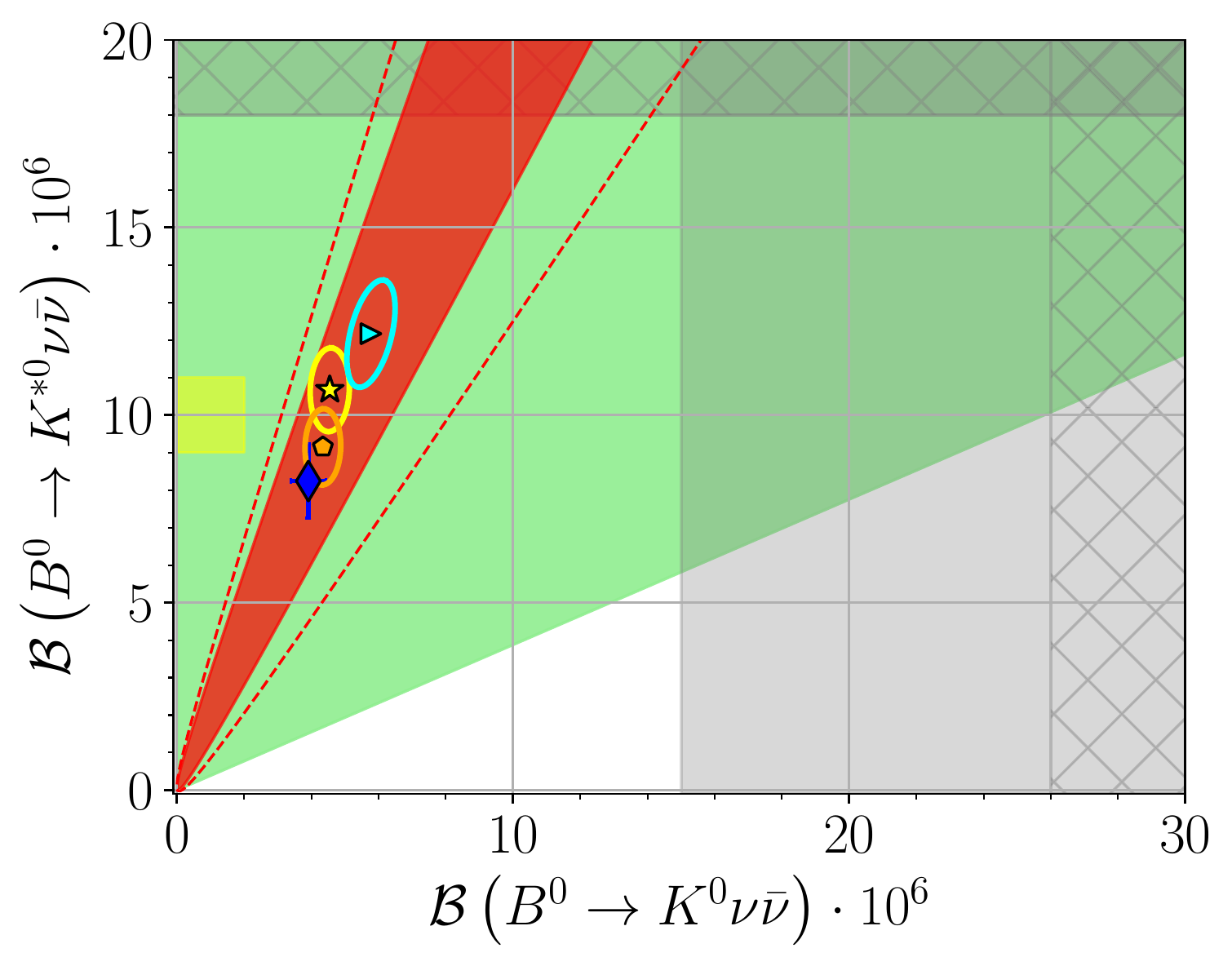}
   \caption{Correlation between $\mathcal{B}(B^0 \to K^{*0}\nu\bar{\nu})$ and $\mathcal{B}(B^0 \to K^0\nu\bar{\nu})$. Details are given in the main text. Figure taken from Ref.~\cite{Bause:2021ply}.
    }
    \label{fig:plotKstarversusK}
\end{figure}

In this section we study $b\to s\,\nu\bar\nu$ transitions and their interplay with $b\to s\,\ell^+\ell^-$ transitions routed by \eqref{eq:super}. The branching ratios for $B\to V\,\nu\bar\nu$ and $B\to P\,\nu\bar\nu$ decays in the LU limit are given by
\begin{align}\label{eq:LUBtoV}
    &\mathcal{B}(B\to V\,\nu\bar\nu)_{\text{LU}}\,=\,A_+^{BV}\,x_{bs,\text{LU}}^+ +\,A_-^{BV}\,x_{bs,\text{LU}}^-~,\quad\mathcal{B}(B\to P\,\nu\bar\nu)_{\text{LU}}\,=\,A_+^{BP}\,x_{bs,\text{LU}}^+~,
\end{align}
where $x_{bs,\text{LU}}^\pm\,=\,3\,\left|\mathcal{C}_{\text{SM}}^{bs\ell\ell}+\mathcal{K}_{L}^{tc\ell\ell}\pm\mathcal{K}_{R}^{bs\ell\ell}\right|^2$, and the values of $A_\pm^{BV}$ and $A_+^{BP}$ for different modes can be found in Ref.~\cite{Bause:2021ply}. We obtain two solutions for the coupling $\mathcal{K}_{L}^{tc\ell\ell}$ when we solve $\mathcal{B}(B\to P\,\nu\bar\nu)_{\text{LU}}$ in Eq.~\eqref{eq:LUBtoV}. Plugging them into Eq.~\eqref{eq:LUBtoV} results in a correlation between both LU branching ratios~\cite{Bause:2021ply}
\begin{align}\label{eq:luregion}
    &\mathcal{B}(B\to V\,\nu\bar\nu)_{\text{LU}}\,=\,\frac{A_+^{BV}}{A_+^{BP}}\,\mathcal{B}(B\to P\,\nu\bar\nu)_{\text{LU}}\,+\,3\,A_{-}^{BV}\,\left|\sqrt{ \frac{\mathcal{B}(B\to P\,\nu\bar\nu)_{\text{LU}}}{3\, A_+^{BP} }}\mp 2\,\mathcal{K}_{R}^{bs\ell\ell}\right|^2~.
\end{align}
The most stringent limits on $\mathcal{K}_{R}^{bs\ell\ell}$ are given for $\ell\ell=\mu\mu$. Performing a 6D global fit of the semileptonic Wilson coefficients $\mathcal{C}_{(7,9,10),\mu}^{(\prime)}$ to the current experimental data on $b\to s\,\mu^+\mu^-$ data (excluding $R_{K^{(*)}}$ which can be polluted by NP effects in electron couplings), we obtain the following $1\sigma$ fit value~\cite{Bause:2021ply}
\begin{align}
  \mathcal{K}_{R}^{bs\ell\ell}&=V_{tb}V^{\ast}_{ts}\,(0.46\pm 0.26)\,.\label{eq:KRglobalfit1sigma_bs}
\end{align}
Fig.~\ref{fig:plotKstarversusK} displays the correlation between $\mathcal{B}(B^0 \to K^{*0}\nu\bar{\nu})$ and $\mathcal{B}(B^0 \to K^0\nu\bar{\nu})$, cf. Eq.~\eqref{eq:luregion}. The SM predictions $\mathcal{B}(B^0 \to K^{*0}\nu\bar{\nu})_{\text{SM}}=(8.2\,\pm\,1.0)\cdot 10^{-6}$, $\mathcal{B}(B^0\to K^0\nu\bar{\nu})_{\text{SM}}=(3.9\,\pm\,0.5)\cdot 10^{-6}$~\cite{Bause:2021ply} are depicted as a blue diamond with their $1\,\sigma$ uncertainties (blue bars). 
We have scanned $\mathcal{K}_{R}^{bs\mu\mu}$, $A_\pm^{B^0 K^{*0}}$, and $A_+^{B^0 K^{0}}$ within their $1\,\sigma$ ($2\,\sigma$) regions in Eq.~\eqref{eq:LUBtoV}, resulting in the dark red region (dashed red lines) which represents the LU region, numerically~\cite{Bause:2021ply}
\begin{align}\label{eq:cone}
\frac{\mathcal{B}(B^0 \to K^{*0}\nu\bar{\nu})}{\mathcal{B}(B^0 \to K^{0}\nu\bar{\nu})}=1.7\ldots2.6 \quad (1.3\ldots 2.9)\,.  
\end{align}
Interestingly, a branching ratio measurement outside the red region would clearly signal evidence for LU violation, but if a future measurement is instead inside this region, this may not necessarily imply LU conservation. Outside the light green region the validity of our effective field theory (EFT) framework gets broken~\cite{Bause:2021ply}. More stringent limits for specific LU SM extensions are depicted as benchmarks, resulting in best fit values (markers) and $1\,\sigma$ regions (ellipses) for $Z^\prime$ (red star), LQ representations $S_3$ (pink pentagon) and $V_3$ (celeste triangle) from $b \to s \,\mu^+ \mu^-$ global fits, see Ref.~\cite{Bause:2021ply} for details.
The current experimental $90\,\%$ CL upper limits, $\mathcal{B}(B^0 \to K^{*0}\nu\bar{\nu})_{\text{exp}}\,<\,1.8\,\cdot\, 10^{-5}$~\cite{Zyla:2020zbs} and $\mathcal{B}(B^0 \to K^0\nu\bar{\nu})_{\text{exp}}\,<\,2.6\,\cdot\, 10^{-5}$~\cite{Zyla:2020zbs}, are displayed by hatched bands. 
The gray bands represent the derived EFT limits, $\mathcal{B}(B^0 \to K^0\nu\bar{\nu})_{\text{derived}}\,<\,1.5\,\cdot\, 10^{-5}$, from Ref.~\cite{Bause:2021ply}. 
A measurement between gray and hatched area would infer a clear hint of NP not covered by our EFT framework, {\it i.e. light particles}. The projected experimental sensitivity ($10\,\%$ at the chosen point) of Belle II with $50\,\text{ab}^{-1}$ is illustrated by the yellow boxes~\cite{Kou:2018nap}. Similar conclusions can be drawn in $b\to d\,\nu\bar\nu$ decay~\cite{Bause:2021ply}.

\section{Conclusions}\label{sec:con}

$SU(2)_L$-invariance relates dineutrinos $\bar q\, q^\prime\, \bar\nu\,\nu$ and charged dilepton couplings $\bar q\,q^\prime\,\ell^+\,\ell^-$ in a model-independent way. This link~\eqref{eq:super} allows probing lepton flavor structure in dineutrino observables in three benchmarks: lepton universality, charged lepton flavor conservation and lepton flavor violation. The link has been exploited for the rare charm and $B$ decays, resulting in novel tests of the aforementioned symmetries, see Table~\ref{tab:afactors} and Eq.~\eqref{eq:cone}, respectively. Our predictions are well-suited for the experiments Belle II~\cite{Kou:2018nap}, BES III~\cite{Ablikim:2019hff}, and future $e^+ e^-$-colliders, such as an FCC-ee running at the $Z$ \cite{Abada:2019lih}, and could offer some insight on the persistent anomalies in $B$ decays.

\section*{Acknowlegments}
We want to thank the organizers for their effort to make this conference such a successful event. This work is supported by the 
\textit{Studienstiftung des Deutschen Volkes} (MG) and the \textit{Bundesministerium f\"ur Bildung und Forschung} -- BMBF (HG).

\end{document}